\documentclass[preprint,aps,superscriptaddress]{revtex4}
\usepackage[dvipdfmx]{graphicx}
\usepackage{dcolumn}
\usepackage{bm}

\begin{document}

\preprint{APS}

\title{Domain dependent Fermi arcs observed in a striped phase dichalcogenide}

\author{T.~Mizokawa}
\affiliation{Department of Applied Physics, 
Waseda University, Shinjuku-ku, Tokyo 169-8555, Japan}
\author{A.~Barinov}
\affiliation{Sincrotrone Trieste S.C.p.A., Area Science Park, 34012 Basovizza, Trieste, Italy}
\author{V.~Kandyba}
\affiliation{Sincrotrone Trieste S.C.p.A., Area Science Park, 34012 Basovizza, Trieste, Italy}
\author{A.~Giampietri}
\affiliation{Sincrotrone Trieste S.C.p.A., Area Science Park, 34012 Basovizza, Trieste, Italy}
\author{R.~Matsumoto}
\affiliation{Department of Applied Physics, 
Waseda University, Shinjuku-ku, Tokyo 169-8555, Japan}
\author{Y.~Okamoto}
\affiliation{Department of Applied Physics, 
Waseda University, Shinjuku-ku, Tokyo 169-8555, Japan}
\author{K.~Takubo}
\affiliation{Department of Chemistry, Tokyo Institute of Technology, Meguro, Tokyo 152-8551, Japan}
\author{K.~Miyamoto}
\affiliation{Hiroshima Synchrotron Radiation Center, Hiroshima University, Hiroshima 739-0046, Japan}
\author{T.~Okuda}
\affiliation{Hiroshima Synchrotron Radiation Center, Hiroshima University, Hiroshima 739-0046, Japan}
\author{S.~Pyon}
\affiliation{Department of Applied Physics, The University of Tokyo, Tokyo 113?8656, Japan}
\author{H.~Ishii}
\affiliation{Department of Physics, Okayama University, Kita-ku, Okayama 700-8530, Japan}
\author{K.~Kudo}
\affiliation{Department of Physics, Osaka University, Toyonaka, Osaka 560-0043, Japan}
\author{M.~Nohara}
\affiliation{Department of Quantum Matter, Hiroshima University, Hiroshima 739-8530, Japan}
\author{N.~L.~Saini}
\affiliation{Department of Physics, Universit\'a di Roma "La Sapienza", 00185 Rome, Italy}

\date{\today}

\begin{abstract}
Angle-resolved photoemission spectromicroscopy on IrTe$_2$ reveals evolution of mesoscopic striped domains across its first order phase transition at $\sim$ 280 K. The striped texture of the domains is characterized by a herringbone arrangement of the electronic anisotropy axes. Under further cooling down to 47 K, the striped domains evolve into trijunction domains with the electronic anisotropy in three directions. Each domain harbors quasi one-dimensional surface bands forming Fermi arcs with peculiar spin polarization. The Fermi arc corresponds to an edge state of the two-dimensional bulk electronic states truncated at the surface, indicating an interesting interplay between the symmetry breaking and the surface electronic states.
\end{abstract}

\maketitle

Domain structures in bistable or polystable electronic systems exhibit various topological shapes that have been one of the subjects of frontier research in fundamental physical science. Some of the fascinating examples are the multiferroic domains with topological nature in YMnO$_3$ \cite{Choi2010} studied by means of electron microscopy and the magnetic domains in perpendicularly magnetized thin films seen by photoemission electron microscopy (PEEM) using circularly polarized light \cite{Kronseder2013}. More recently, novel domain textures of antiferromagnets are observed in Nd$_2$Ir$_2$O$_7$ by means of scanning microwave impedance microscopy \cite{Ma2015} and in Fe$_2$O$_3$ by means of photoelectron microscopy \cite{Chmiel2018}. Metallic and insulating domains in systems with Mott transitions and colossal magnetoresistance were also imaged by photoemission spectromicroscopy, revealing that their shapes are not associated with the underlying lattice direction and the symmetry breaking \cite{Lupi2010,Sarma2004}. In general, mesoscopic domain structures of symmetry-broken electronic systems are not directly related to the nature of the broken symmetry itself, but determined by the interfacial energy minimum. A rare case is the shape memory alloy NiMn$_2$Ga in which the symmetry-broken electronic state is associated with the domain texture and the shape memory effect \cite{Jenkins2012}. 

Among various systems showing first-order phase transitions, CuIr$_2$S$_4$ with Ir pyrochlore lattice shows an interesting transition with octamer charge order \cite{Radaelli2002}, which is accompanied by a dramatic Ir 5d orbital symmetry breaking \cite{Khomskii2005}. The layered IrTe$_2$ with triangular lattice of Ir exhibits a structural phase transition from trigonal to monoclinic\cite{Jobic1991,Matsumoto1999}, similar to that of CuIr$_2$S$_4$, at about 280 K. Since the layered IrTe$_2$ crystals can be cleaved to obtain flat surfaces, IrTe$_2$ is best suited for photoemission studies unlike CuIr$_2$S$_4$. The single crystal samples of IrTe$_2$ reveal two structural phase transitions around 280 K and 180 K\cite{Pyon2013}. In addition, interesting interplay between the structural transition and superconductivity was found in IrTe$_2$ \cite{Yang2012,Pyon2012} in which multi-band Fermi surfaces are expected to play a significant role \cite{Ootsuki2012}. An optical study has shown that the band structure is reconstructed over a broad energy scale up to ~2 eV \cite{Fang2013} indicating importance of the Te 5p orbitals together with the Ir 5d orbitals \cite{Oh2013,Cao2013}. The interplay between the Ir 5d and Te 5p orbitals is essential that has also been indicated by extended x-ray absorption fine structure, photoemission, and high pressure studies \cite{Joseph2013,Ootsuki2013,Kim2015,Kiswandhi2013}. The anisotropic charge/orbital order in IrTe$_2$ was identified by x-ray diffraction and ab-initio calculations \cite{Toriyama2014,Pascut2014}, and further indicated in scanning tunneling microscopy studies \cite{Ko2015,Hsu2013,Machida2013,Kim2016}. More recently, complex phase evolution due to a heterogeneous nucleation mechanism has been revealed by scanning Raman microscopy \cite{Oike2021}. Such anomalous phase evolution would be related to the anisotropic charge/orbital order. Since the band structure is sensitive to the charge/orbital order, angle-resolved photoemission spectromicroscopy is a unique technique to study if such an order occurs in domains, and if so, what is the domain dependent electronic structure of IrTe$_2$. Here, it is worth noting that very recently the technique has been successfully applied to study facet dependent surface states in weak \cite{Noguchi2019} or higher order \cite{Noguchi2021} topological insulators.

Single crystal samples of IrTe$_2$ were prepared using a self-flux as reported in the literature \cite{Pyon2013}. The photoemission microscopy results were obtained at the spectromicroscopy beamline, Elettra synchrotron facility, Italy \cite{Dudin2010}. Photons at 27 eV were focused through a Schwarzschild objective in order to obtain a submicron size beam spot. For the present measurements, the total energy resolution was set to about 50 meV and the angle resolution was 1 degree. The measurements were performed under ultra-high vacuum in the 10$^{-10}$ Torr range on in-situ prepared surfaces. A standard photoemission microscopy procedure was used to remove topographic features from the images presented \cite{Marsi1997}. The microscopy images were obtained by integrating the photoemission spectra in appropriate energy windows. The photoemission spectra are angle integrated with the acceptance angle of $\pm7.5$ degrees corresponding to $\pm0.3 \AA^{-1}$ around the $\Gamma$ point along the $\Gamma$M direction. Spin-resolved photoemission spectroscopy (SARPES) measurements were performed at beamline 9B of Hiroshima Synchrotron Radiation Center (HiSOR). The base pressure of the spectrometer was in the 10$^{-11}$ Torr range. The crystals were cleaved at 300 K under the ultrahigh vacuum and then cooled to 20 K for SARPES. The total energy resolution was set to 51 meV for excitation energy h$\nu$ of 23 eV. 

Figure 1(a) shows the photoemission spectromicroscopy image of the entire single crystal of IrTe$_2$ taken at 250 K. The images were obtained by integrating the photoemission spectra from the Fermi level (${\rm E_F}$) to -2.3 eV since the spectrum of IrTe$_2$ in this energy range is known to change dramatically across the phase transition. The crystal was cleaved at 300 K and slowly cooled across the phase transition temperature of 280 K. The zoomed view is shown in Fig. 1(b). The image shows striped domains of length $\sim$ 50-100 $\mu$m. The intensity of each pixel reflects the spectral change due to the first order phase transition and the high (bright) and low (dark) intensity regions, may correspond to the different directions of the charge/orbital arrangements. Figures 1(c)-1(e) exhibit temperature evolution of the striped domains with cooling. Here, the cooling rate is set to $\sim$ 2 K/min that is slow enough to allow formation of low temperature phase. Apparently, the striped domains contrast is weak at 47 K. The image taken at this temperature suggests complicated domain structure of low temperature phase which will be discussed in the following paragraphs. After the observation, the sample was warmed to 290 K where the spectromicroscopy image hardly shows any appreciable structures [Fig. 1(f)]. It is expected that each domain is characterized by an anisotropic electronic state driven by the anisotropic charge/orbital order with Ir$^{4+}$-like dimerized sites forming stripes as illustrated in Fig. 1(g). In addition, the dispersive direction of the quasi one-dimensional surface bands corresponds to the direction of the charge/orbital stripes, and the Fermi arcs created by the surface bands are perpendicular to the stripes as shown in Figs. 1(g) and 1(h) \cite{Ootsuki2017}.

Figure 2 shows typical photoemission spectra of selected domains at 290 K, 250 K, and 47 K. At 290 K above the transition, the system is homogeneous with a single photoemission spectrum [Figs. 2(a) and 2(b)]. In the striped domains at 250 K, only two types of photoemission spectra are seen although three different directions of the anisotropic charge/orbital order are expected in the triangular lattice [Figs. 2(c) and 2(d)]. Indeed at 47 K, three types of photoemission spectra are observed, consistent with the three directions of the charge/orbital order [Figs. 2(e) and 2(f)]. The lower temperature bulk phase transition at 180 K is accompanied by periodicity change of the charge/orbital order from $5\times1\times5$ to $8\times1\times8$ with cooing \cite{Ko2015,Hsu2013}. In addition, it has been reported that the surface structure changes from high temperature $5\times1$ phase to low temperature $8\times1$ phase with contribution of $6\times1$ phase in the intermediate temperature range \cite{Rumo2020}. The periodicity change would drive the domains structure to evolve dramatically.

Figures 3(a) and 3(b) show the photoemission intensity images at 47 K for the two different energy windows indicated by the thick arrows in Fig. 2(f). These images allow us to draw boundaries of the three kinds of domains as shown by solid lines in Fig. 3(c). Figures 3(d), 3(e), and 3(f) show Fermi surfaces measured at the three points [indicated by the crosses in Fig. 3(c)] belonging to different domains. The electronic anisotropy can be recognized by their shapes. In addition to the Fermi surfaces from the bulk, Fermi arcs derived from the surface bands are observed for each domain. The surface bands in domain \#1 are dispersive along the direction of the charge/orbital stripes and exhibit the Fermi arcs perpendicular to them as indicated by the broken lines in Fig. 3(d). Compared with the sharp Fermi arcs for domain \#1, the Fermi arcs of domains \#2 and \#3 are blurred although their anisotropic shape can be recognized as indicated by the broken lines in Figs. 3(e) and 3(f). Most likely, the charge/orbital stripes are well established in newly developed domain \#1 whereas they are somewhat disordered in domains \#2 and \#3 during the evolution from the intermediate temperature 5-fold stripe phase to the low temperature 6- or 8-fold stripe phase \cite{Ko2015}. The striped domain pattern in the intermediate temperature range is characterized by the two directions of electronic stripes arranged in the herringbone structure as shown in Fig. 3(g). Such a herringbone arrangement can be expected in the electronic stripe phase (or nematic phase) in a square lattice. On the other hand, in the triangular lattice, the herringbone arrangement with striped domain is surprising. The domain structure is expected to depend on the cooling rate as shown in Fig. S1(a) and S1(b) in the Supplemental Materials displaying photoemission intensity images for the samples quenched quickly to 47 K after cleavage at 300 K and cleaved at 47 K after quenching from 300 K, respectively. The major domains have irregular shape and are much larger in size than those seen for the slow cooling. In the quenched cases, one of the three directions is selected and frozen at the phase transition. The present observation suggests that the anisotropic surface state can be controlled and manipulated by domain engineering in IrTe$_2$.

The Fermi arcs are mainly constructed from the Ir 5d and Te 5p orbitals at the surface layer. Therefore, the Fermi arcs may exhibit spin polarization induced by broken inversion symmetry and strong spin-orbit interaction of Ir 5d and Te 5p. Indeed, the Fermi arcs of the striped phase have a peculiar spin texture. Figure 4 shows spin-resolved angle-resolved photoemission spectroscopy (SARPES) spectra for out-of-plane spin polarization taken at 20 K with linearly polarized light along a cut crossing the Fermi arc. Since the crystal was cleaved at 300 K and then rapidly cooled to 20 K with cooling rate of 20-30 K/min, the situation is close to the quenched case shown in Fig. S1(a) in the Supplemental Materials. The sample surface is dominated by the domain with horizontal stripes in Fig. S1(a), and the Fermi arcs along the vertical direction are clearly observed as shown in Fig. S1(c). Similarly in Fig. 4, the sample surface is dominated by the horizontal stripes, and the surface band creating the vertical Fermi arc can be observed at 20 K as shown in Fig. 4(b) with the relatively large beam spot for SARPES. Here, $k_x$ ($k_y$) represents the momentum perpendicular (parallel) to the Fermi arc. The momentum positions for the SARPES spectra in Fig. 4(c) are indicated by open triangles and circles, respectively, in Fig. 4(a). The out-of-plane spin polarization is also plotted in Fig. 4(c). While the bulk bands around -1.3 eV and -2.3 eV do not show appreciable spin polarization, the surface bands near ${\rm E_F}$ and at -0.5 eV exhibit negative spin polarization for the positive $k_x$ (Points h5-h7). The spin polarization of the surface bands is clearly seen in the upper panel of Fig. 4(d), in which the out-of-plane spin polarization is plotted as a function of $k_x$ and energy. According to the calculations by Pascut et al. and Toriyama et al., the charge-orbital stripe phase is characterized by new conducting layers which are tilted from the IrTe$_2$ plane \cite{Pascut2014,Toriyama2014}. Therefore, the spin direction at the edge of new conducting layers can be tilted with respect to the original IrTe$_2$ plane. In the lower panel of Fig. 4(d), the out-of-plane spin polarization is plotted as a function of $k_y$ and energy. Interestingly, the out-of-plane spin polarization changes its sign along the Fermi arc while the in-plane spin polarization does not as shown in Figs. S2 and S3 in the Supplemental Materials.

In summary, using angle-resolved photoemission spectromicroscopy we have shown that the stripe-type charge/orbital order in IrTe$_2$ is accompanied by mesoscopic striped domains. The striped domain texture is stabilized in the herringbone arrangement of the charge/orbital stripes between 250 K and 120 K. The striped domain of the electronic stripe phase is formed under anisotropic strain from the neighboring domains similar to the one in shape memory alloys. At 47 K, the striped texture is replaced by the trijunction texture of the three types of domains with different charge/orbital stripe directions. On the other hand, formation of the striped or trijunction texture is suppressed in the quenched case. Each domain harbors anisotropic surface states forming Fermi arcs with peculiar spin polarization. The Fermi arcs correspond to the edge states of the two-dimensional bulk electronic states with the charge/orbital stripe which are truncated at the cleaved surface. IrTe$_2$ is very unique in that the spin polarized surface states can be controlled by the domain texture of charge/orbital stripe order. We anticipate that the striped domain boundary in IrTe$_2$ is a new playground to explore a novel low-dimensional electronic state under the strong spin-orbit interaction with a possible impact on nanoscale electronic/magnetic devices.

The authors would like to thank D. Ootsuki and M. Horio for contributions to the experiments in the early stage of this work. The authors also thank J. Jia for assisting during a part of the experiments. This work was partially supported by Grants-in-Aid from the Japan Society of the Promotion of Science (JSPS) (No. 25400372, 26287082, 26400349, 25400356). This work was supported by joint research program of ZAIKEN, Waseda University (Project No.31010).

\clearpage
Figure captions:\\
Figure 1:\\
(color online) 
(a) Photoemission intensity image of the wide area at 250 K. (b) Photoemission intensity image of the selected area indicated by the box in (a) at 250 K. (c-e) Photoemission intensity images of the selected area at 220 K, 120 K, and 47 K taken in the cooling process. The cooling rate is $\sim$ 2 K/min. (f) Photoemission intensity image of the selected area at 290 K after the heating from 47 K. All the images were taken at h$\nu$=27 eV with linear polarization (horizontal). (g) Three directions of the charge/orbital stripes in the IrTe$_2$ layer. (h) Three directions of the Fermi arcs in the Brillouin zone corresponding to the three directions of the charge/orbital stripes. The dotted curves represent the bulk Fermi surfaces.
\\
\\
Figure 2:\\
(color online) 
(a) Domain structure and (b) photoemission spectra at 290 K. (c) Domain structure and (d) photoemission spectra at 250 K. (e) Domain structure and (f) photoemission spectra at 47 K. The images and spectra were taken at h$\nu$=27 eV with linear polarization (horizontal).
\\
\\
Figure 3:\\
(color online) 
(a-c) Domain textures at 47 K visualized by different energy windows. (d-f) Fermi surfaces of the three different domains at 47 K. The broken lines indicate Fermi arcs derived from the surface states (SS). (g-i) Domain textures at 250 K, 120 K, and 47 K. All the images were taken at h$\nu$=27 eV with linear polarization (horizontal). The solid lines indicate domain boundaries. The arrows indicate the directions of charge/orbital stripes.
\\
\\
Figure 4:\\
(color online) 
(a) Fermi surface map taken at 300 K. The solid curves indicate the calculated Fermi surfaces taken from the literature \cite{Ootsuki2013}. Momentum points for SARPES are indicated by open triangles and circles. (b) Band maps along the horizontal (A-H) cut at 300 K and 20 K. At 20 K, the surface bands cross ${\rm E_F}$ as indicated by the triangles. (c) SARPES spectra at selected points across and along the Fermi arc taken by linearly polarized h$\nu$ = 23 eV at 20 K. Spin polarization direction is perpendicular to the surface. (d) Spin polarization maps across and along the Fermi arc. Positive and negative spin polarizations are indicated by red and blue colors. The broken lines labeled as SS indicate the surface band.


\begin{thebibliography}{99}

\bibitem{Choi2010}
T. Choi, Y. Horibe, H. T. Yi, Y. J. Choi, W. Wu, and S.-W. Cheong, Nature Mater. {\bf 9}, 253 (2010).

\bibitem{Kronseder2013}
M. Kronseder, M. Buchner, H. G. Bauer, and C. H. Back, Nature Commun. {\bf 4}, 2054 (2013).

\bibitem{Ma2015}
E. Y. Ma, Y.-T. Cui, K. Ueda, S. Tang, K. Chen, N. Tamura, P. M. Wu, J. Fujioka, Y. Tokura, Z.-X. Shen, Science {\bf 350}, 538 (2015).

\bibitem{Chmiel2018}
F. P. Chmiel, N. W. Price, R. D. Johnson, A. D. Lamirand,  J. Schad, G. van der Laan, D. T. Harris, J. Irwin, M. S. Rzchowski, C.-B. Eom, and P. G. Radaelli, Nature Mater. {\bf 17}, 581 (2018).

\bibitem{Lupi2010}
5. S. Lupi, L. Baldassarre, B. Mansart, A. Perucchi, A. Barinov, P. Dudin, E. Papalazarou, F. Rodolakis, J. -P. Rueff, J. -P. Iti?, S. Ravy, D. Nicoletti, P. Postorino, P. Hansmann, N. Parragh, A. Toschi, T. Saha-Dasgupta, O. K. Andersen, G. Sangiovanni, K. Held, and M. Marsi, Nature Commun. {\bf 1}, 105 (2010).

\bibitem{Sarma2004}
D. D. Sarma, D. Topwal, U. Manju, S. R. Krishnakumar, M. Bertolo, S. La Rosa, G. Cautero, T. Y. Koo, P. A. Sharma, and A. Fujimori, Phys. Rev. Lett. {\bf 93}, 097202 (2004).

\bibitem{Jenkins2012}
C. A. Jenkins, A. Scholl, R. Kainuma, H. J. Elmers, and T. Omori, Appl. Phys. Lett. {\bf 100}, 032401 (2012).

\bibitem{Radaelli2002}
P. G. Radaelli, Y. Horibe, M. J. Gutmann, H. Ishibashi, C. H. Chen, R. M. Ibberson, Y. Koyama, Y. S. Hor, V. Kirykhin, and S. W. Cheong, Nature {\bf 416}, 155 (2002).

\bibitem{Khomskii2005}
D. I. Khomskii and T. Mizokawa, Phys. Rev. Lett. {\bf 94}, 156402 (2005).

\bibitem{Jobic1991}
S. Jobic, P. Deniard, R. Brec, J. Rouxel, A. Jouanneaux, and A. N. Fitch, Z. Anorg. Allg. Chem. {\bf 598}, 199 (1991).

\bibitem{Matsumoto1999}
N. Matsumoto, K. Taniguchi, R. Endoh, H. Takano, and S. Nagata, J. Low Temp. Phys. {\bf 117}, 1129 (1999).

\bibitem{Pyon2013}
S. Pyon, K. Kudo, and M. Nohara, Physica C {\bf 494}, 80 (2013).

\bibitem{Pyon2012}
S. Pyon, K. Kudo, and M. Nohara, J. Phys. Soc. Jpn. {\bf 81}, 053701 (2012).

\bibitem{Yang2012}
J. J. Yang, Y. J. Choi, Y. S. Oh, A. Hogan, Y. Horibe, K. Kim, B. I. Min, and S-W. Cheong, Phys. Rev. Lett. {\bf 108}, 116402 (2012).

\bibitem{Ootsuki2012}
D. Ootsuki, Y. Wakisaka, S. Pyon, K. Kudo, M. Nohara, M. Arita, H. Anzai, H. Namatame, M. Taniguchi, N. L. Saini, and T. Mizokawa, Phys. Rev. B {\bf 86}, 014519 (2012).

\bibitem{Fang2013}
A. F. Fang, G. Xu, T. Dong, P. Zheng, and N. L. Wang, Sci. Rep. {\bf 3}, 1153 (2013).

\bibitem{Oh2013}
Y.S. Oh, J. J. Yang, Y. Horibe, and S-W. Cheong, Phys. Rev. Lett. {\bf 110}, 127209 (2013).

\bibitem{Cao2013}
H. Cao, B. C. Chakoumakos, X. Chen, J. Yan, M. A. McGuire, H. Yang, R. Custelcean, H. D.  Zhou, D. J. Singh, and D. Mandrus, Phys. Rev. B {\bf 88}, 115122 (2013).

\bibitem{Joseph2013}
B. Joseph, M. Bendele, L. Simonelli, L. Maugeri, S. Pyon, K. Kudo, M. Nohara, T. Mizokawa, and N. L. Saini, Phys. Rev. B {\bf 88}, 224109 (2013).

\bibitem{Ootsuki2013}
D. Ootsuki, S. Pyon, K. Kudo, M. Nohara, M. Arita, H. Anzai, H. Namatame, M. Taniguchi, N. L. Saini, and T. Mizokawa, J. Phys. Soc. Jpn. {\bf 82}, 093704 (2013).

\bibitem{Kim2015}
K. Kim, S. Kim, K.-T. Ko, H. Lee, J.-H. Park, J. J. Yang, S-W. Cheong, and B. I. Min, Phys. Rev. Lett. {\bf 114}, 136401 (2015).

\bibitem{Kiswandhi2013}
A. Kiswandhi, J. S. Brooks, H. B. Cao, J. Q. Yan, D. Mandrus, Z. Jiang, and H. D. Zhou, Phys. Rev. B {\bf 87}, 121107(R) (2013).

\bibitem{Toriyama2014}
T. Toriyama, M. Kobori, Y. Ohta, T. Konishi, S. Pyon, K. Kudo, M. Nohara, K. Sugimoto, T. Kim, and A. Fujiwara, J. Phys. Soc. Jpn. {\bf 83}, 033701 (2014).

\bibitem{Pascut2014}
G. L. Pascut, K. Haule, M. J. Gutmann, S. A. Barnett, A. Bombardi, S. Artyukhin, T. Birol, D. Vanderbilt, J. J. Yang, S-W. Cheong, and V. Kiryukhin, Phys. Rev. Lett. {\bf 112}, 086402 (2014).

\bibitem{Ko2015}
K.-T. Ko, H.-H. Lee, D.-H. Kim, J. J. Yang, S-W. Cheong, M. J. Eom, J. S. Kim, R. Gammag, K.-S. Kim, H.-S. Kim, T.-H. Kim., H. W. Yeom, T.-Y. Koo, H.-D. Kim, and J.-H. Park, Nature Commun. {\bf 6}, 7342 (2015).

\bibitem{Hsu2013}
P.-J. Hsu, T. Mauerer, M. Vogt, J. J. Yang, Y. S. Oh, S-W. Cheong, M. Bode, and W. Wu, Phys. Rev. Lett. {\bf 111}, 266401 (2013).

\bibitem{Machida2013}
T. Machida, Y. Fujisawa, K. Igarashi, A. Kaneko, S. Ooi, T. Mochiku, M. Tachiki, K. Komori, K. Hirata, and H. Sakata, Phys. Rev. B {\bf 88}, 245125 (2013).

\bibitem{Kim2016}
H. S. Kim, S. Kim, K. Kim, B. I. Min, Y-H. Cho, L. Wang, S-W. Cheong, and H. W. Yeom, Nano Lett. {\bf 16}, 4260 (2016).

\bibitem{Oike2021}
H. Oike, K. Takeda, M. Kamitani, Y. Tokura, and F. Kagawa, Phys. Rev. Lett. {\bf 127} 145701 (2021).

\bibitem{Noguchi2019}
R. Noguchi, T. Takahashi, K. Kuroda, M. Ochi, T. Shirasawa, M. Sakano, C. Bareille, M. Nakayama, M. D. Watson, K. Yaji, A. Harasawa, H. Iwasawa, P. Dudin, T. K. Kim, M. Hoesch, V. Kandyba, A. Giampietri, A. Barinov, S. Shin, R. Arita, T. Sasagawa, and T. Kondo, Nature {\bf 566}, 518 (2019).

\bibitem{Noguchi2021}
R. Noguchi, M. Kobayashi, Z. Jiang, K. Kuroda, T. Takahashi, Z. Xu, D. Lee, M. Hirayama, M. Ochi, T. Shirasawa, P. Zhang, C. Lin, C. Bareille, S. Sakuragi, H. Takaka, S. Kunisada, K. Kurokawa, K. Yaji, A. Harasawa, V. Kandyba, A. Giampietri, A. Barinov, T. K. Kim, C. Cacho, M. Hashimoto, D. Lu, S. Shin, R. Arita, K. R. Lai, T. Sasagawa, and T. Kondo, Nat. Mater. {\bf 20}, 473 (2021).

\bibitem{Dudin2010}
P. Dudin, P. Lacovig, C. Fava, E. Nicolini, A. Bianco, G. Cautero, and A. Barinov, J. Synchrotron Rad. {\bf 17}, 445-450 (2010).

\bibitem{Marsi1997}
M. Marsi, L. Casalis, L. Gregoratti, S. Gunther, A. Kolmakov, J. Kovac, D. Lonza, and M. Kiskinova, J. Electron Spectrosc. Relat. Phenom. {\bf 84}, 73-83 (1997).

\bibitem{Ootsuki2017}
D. Ootsuki, H. Ishii, K. Kudo., M. Nohara, M. Takahashi, M. Horio, A. Fujimori, T. Yoshida, M. Arita, H. Anzai, H. Namatame, M. Taniguchi, N. L. Saini, and T. Mizokawa, J. Phys. Sco. Jpn. {\bf 86} 123704 (2017).

\bibitem{Rumo2020}
M. Rumo, C. W. Nicholson, A. Pulkkinen, B. Hildebrand, G. Kremer, B. Salzmann, M.-L. Mottas, K. Y. Ma, E. L. Wong, M. K. L. Man, K. M. Dani, B. Barbiellini, M. Muntwiler, T. Jaouen, F. O. von Rohr, and C. Monney, Phys. Rev. B {\bf 101} 235120 (2020).

\end{thebibliography}
\end{document}